\documentclass[bibyear]{aa}
\usepackage{graphicx}
\usepackage{float}
\usepackage{mathrsfs}
\newcommand{\ti}{$^{44}$Ti}
\newcommand{\sca}{$^{44}$Sc}
\newcommand{\ca}{$^{44}$Ca}
\newcommand{\cxiv}{$^{14}$C}
\newcommand{\bex}{$^{10}$Be}
\newcommand{\Xd}{$\chi^2$}
\newcommand{\rXd}{$\chi^2_\nu$}

\begin{document}

\title{Long-term evolution of the heliospheric magnetic field \\ inferred from cosmogenic \ti\ activity in meteorites}

\author{S. Mancuso\inst{1}, C. Taricco\inst{1,2}, P. Colombetti\inst{2}, S. Rubinetti\inst{2}, N. Sinha\inst{3}, N. Bhandari\inst{4}}    
                                                                            
\institute{Istituto Nazionale di Astrofisica, Osservatorio Astrofisico di Torino, Strada Osservatorio 20, Pino Torinese 10025, Italy \\ 
\email{mancuso@oato.inaf.it}
\and Dipartimento di Fisica, Universit\`a di Torino, Via P. Giuria 1, 10125 Torino, Italy
\and Wentworth Institute of Technology, Boston, MA, USA
\and Physical Research Laboratory and Basic Sciences Research Institute, Navrangpura, Ahmedabad, India}

\date{Received / Accepted}

\abstract{Typical reconstructions of historic heliospheric magnetic field (HMF)  $B_{\rm HMF}$ are based on the analysis of the sunspot activity, geomagnetic data or on measurement of cosmogenic isotopes stored in terrestrial reservoirs like trees (\cxiv) and ice cores (\bex).
The various reconstructions of $B_{\rm HMF}$ are however discordant both in strength and trend.
Cosmogenic isotopes, which are produced by galactic cosmic rays (GCRs) impacting on meteoroids and whose production rate is modulated by the varying HMF convected outward by the solar wind, may offer an alternative tool for the investigation of the HMF in the past centuries. 
In this work, we aim to  evaluate the long-term evolution of $B_{\rm HMF}$ over a period covering the past twenty-two solar cycles by using measurements of the cosmogenic \ti\ activity ($\tau_{1/2} = 59.2 \pm 0.6$ yr) measured in 20 meteorites which fell between 1766 and 2001.
Within the given uncertainties, our result is compatible with a HMF increase from $4.87^{+0.24}_{-0.30}$ nT in 1766 to $6.83^{+0.13}_{-0.11}$ nT in 2001, thus implying an overall average increment of $1.96^{+0.43}_{-0.35}$ nT over 235 years since 1766 reflecting the modern Grand maximum.
The $B_{\rm HMF}$ trend thus obtained is then compared with the most recent reconstructions of the near-Earth heliospheric magnetic field strength based on geomagnetic, sunspot number and cosmogenic isotope data.
}

\titlerunning{Long-term evolution of the heliospheric magnetic field intensity}
\authorrunning{Mancuso et al.}
\maketitle

\section{Introduction}

The heliospheric magnetic field (HMF) in the interplanetary space is the extension of the magnetic field of the sun and its corona that is carried out into space by the highly conductive solar wind. 
Since 1962, when it was first estimated by magnetometers on board satellites, continuous measurements have been made near Earth, providing a record of the HMF in the last five decades (see Owens \& Forsyth 2013 for a comprehensive review).
This record has shown that the near-Earth HMF increases from the minima of solar activity to the maxima, varying considerably from one cycle to the next.
A number of sources of proxy data, such as sunspots, geomagnetic indices and aurorae, allow insight into long-term HMF variability that is supposed to have changed substantially over the past centuries. 
Sunspot records from 1610 to present relate to large-scale magnetic features on the photosphere, thus requiring modeling of the Sun's open magnetic flux to convert sunspot number observations to HMF strength (e.g., Solanki et al. 2000; Wang \& Sheeley 2003).
As for the geomagnetic proxies, although there is good agreement in HMF reconstructions from different geomagnetic records in the past century (Lockwood \& Owens 2011), uncertainties increase prior to about 1880 due to the decrease of the number and quality of geomagnetic station records.
Cosmogenic radionuclide data from \bex\ and \cxiv, which are produced in the atmosphere and deposited, respectively, in ice cores and trees, are also a reliable source of information regarding the HMF in the past (see, e.g., Beer et al. 2012; Usoskin 2017).
Their production rate depends on the intensity of the galactic cosmic rays (GCR) that penetrate into the Earth's atmosphere.
Before they reach the Earth, these GCRs travel through the heliosphere where they are modulated by the varying HMF convected outward by the solar wind with the result that their flux is inversely correlated with the HMF strength (e.g., Potgieter 2013).
Reconstructions of HMF strength in the past by means of the above proxies have been however discordant.
In particular, \bex\ and \cxiv\ concentrations are affected by terrestrial phenomena such as geomagnetic field, climatic changes, deposition rate variations, and exchange within the various terrestrial reservoirs, all of which tend to mask the modulation due to solar activity.
Use of \cxiv\ is moreover made more difficult since about 1850 due to the increasing influence of burning of fossil fuel and the nuclear bomb tests commencing in the early 1950s.

\begin{figure*}
\centering
\includegraphics[width=16cm]{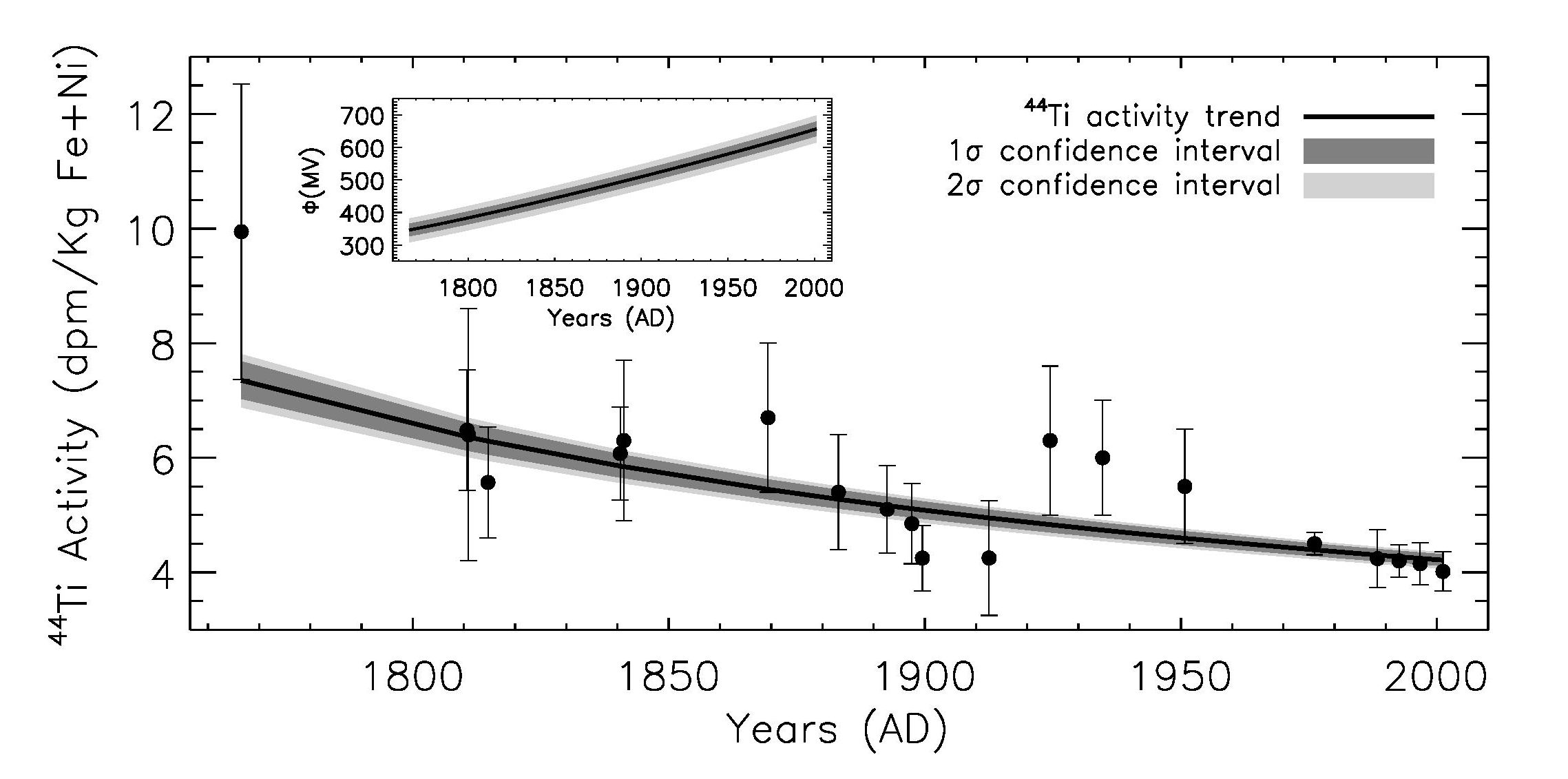}
\caption{
Time profile of the measured \ti\ activity in units of disintegrations per minute per kg of Fe + Ni for meteorites which fell between 1766 and 2001.
Black dots depict measurements in meteorites. 
Error bars correspond to $1\sigma$ uncertainties in the \ti\ activity.
The black curve represents the trend of the \ti\ activity as given by a $\chi^2$ fit of Model 11 to the data.
Dark-gray and light-gray shaded areas denote, respectively, the 1$\sigma$ (68.27\%) and 2$\sigma$ (95.45\%) confidence bands. 
The inset shows the time profile of the modulation potential (in MV) obtained from inverting Eq. (3) (see text for details).
}
\end{figure*}

Cosmogenic radioisotopes in meteorites, which are produced when the rocks are exposed to GCRs in the heliospheric space, offer an alternative tool for the investigation of the HMF variations in the past. 
These nuclides are produced directly in the meteoroid body in space, so that the problematics affecting \bex\ and \cxiv\ are totally bypassed.
The activity of cosmogenic radioisotopes in meteoroids is regulated by an integral of the balance between their production and decay, yielding the time-integrated GCR flux over a period approximately determined by the mean life of the isotopes. 
By measuring the abundance of relatively short-lived cosmogenic radioisotopes in meteorites which fell in the past, it is possible to infer the variability of the GCR flux, since their production ends after the fall of the meteorite.
We previously demonstrated that \ti, with a half-life of $59.2 \pm 0.6$ yr, is an ideal index in meteorites for monitoring solar activity over the last two-to-three centuries, free from interference due to terrestrial processes (Bonino et al. 1995; Taricco et al. 2006; Usoskin et al. 2006). 
This isotope is mainly produced by spallation reactions between GCR protons ($>70$ MeV) with nuclei of Fe and Ni in the body of the meteoroids.

In this work, we show that measurements of the \ti\ activity in meteorites can be used for reconstructing the long-term evolution of the HMF over the last centuries. 
The $B_{\rm HMF}$ trend between 1766 and 2001 will be compared to various reconstructions of the HMF obtained by using different proxies.

\begin{table*}
\caption{Fit quality evaluation criteria for 16 competing model functions}
\centering
\scalebox{0.85}{
\begin{tabular}{ccccccccccccc}
\hline\hline
 Model no. & Fitting function & p& A & B & C & \Xd  &  d.o.f.  & \rXd  & AIC  & $\Delta$AIC  & BIC  & $\Delta$BIC   \\  [0.1em]
\hline 
   1 &                  A+B{\it t} &  2 & 2.535$\cdot 10^1$ &-1.060$\cdot 10^{-2}$ &          &  10.494 & 18  &    0.58  &   -8.19  &    3.50  &   -6.91  &    4.01  \\
   2 &              A+B{\it t}$^2$ &  2 & 1.517$\cdot 10^1$ &-2.755$\cdot 10^{-6}$ &          &  10.555 & 18  &    0.59  &   -8.08  &    3.62  &   -6.79  &    4.13  \\
   3 &             A{\it t}+B{\it t}$^2$ &  2 & 1.579$\cdot 10^{-2}$ &-6.858$\cdot 10^{-6}$ &          &  10.658 & 18  &    0.59  &   -7.88  &    3.81  &   -6.60  &    4.33  \\
   4 &          A+B{\it t}+C{\it t}$^2$ &  3 & 1.101$\cdot 10^2$ &-9.900$\cdot 10^{-2}$ & 2.301$\cdot 10^{-5}$ &  10.254 & 17  &    0.60  &   -5.86  &    5.84  &   -4.37  &    6.55  \\
   5 &             A exp(B{\it t}) &  2 & 3.065$\cdot 10^2$ &-2.148$\cdot 10^{-3}$ &          &  10.280 & 18  &    0.57  &   -8.61  &    3.09  &   -7.32  &    3.60  \\
   6 &           A exp(B{\it t})+C &  3 &-2.784$\cdot 10^1$ & 2.399$\cdot 10^{-4}$ & 4.913$\cdot 10^1$ &  10.520 & 17  &    0.62  &   -5.35  &    6.35  &   -3.86  &    7.06  \\
   7 &         A exp[B/({\it t}+C)] &  3 & 1.184$\cdot 10^2$ &-2.148$\cdot 10^{-3}$ &-4.427$\cdot 10^2$ &  10.280 & 17  &    0.60  &   -5.81  &    5.88  &   -4.32  &    6.60  \\
   8 &             A exp(B/{\it t}) &  2 & 8.106$\cdot 10^{-2}$ & 7.888$\cdot 10^3$ &          &  10.171 & 18  &    0.57  &   -8.82  &    2.88  &   -7.53  &    3.39  \\
   9 &           A exp(B/{\it t})+C &  3 &-1.778$\cdot 10^2$ &-2.508$\cdot 10^2$ & 1.610$\cdot 10^2$ &  10.387 & 17  &    0.61  &   -5.60  &    6.09  &   -4.12  &    6.81  \\
  10 &         A exp[B/({\it t}+C)] &  3 & 2.836$\cdot 10^{0}$ & 1.482$\cdot 10^2$ &-1.635$\cdot 10^3$ &   9.732 & 17  &    0.57  &   -6.91  &    4.79  &   -5.42  &    5.50  \\
  11 &                {\it t}/(A+{\it t}) &  1 &-1.526$\cdot 10^3$ &          &          &   9.973 & 19  &    0.52  &  -11.70  &    0.00  &  -10.92  &    0.00  \\
  12 &              A{\it t}/(B+{\it t}) &  2 & 9.733$\cdot 10^{-1}$ &-1.537$\cdot 10^3$ &          &   9.941 & 18  &    0.55  &   -9.28  &    2.42  &   -7.99  &    2.93  \\
  13 &            A+B{\it t}/(C+{\it t}) &  3 & 2.513$\cdot 10^{0}$ & 2.690$\cdot 10^{-1}$ &-1.691$\cdot 10^3$ &   9.698 & 17  &    0.57  &   -6.98  &    4.72  &   -5.49  &    5.43  \\
  14 &              A+B ln({\it t}) &  2 & 1.590$\cdot 10^2$ &-2.037$\cdot 10^1$ &          &  10.435 & 18  &    0.58  &   -8.30  &    3.39  &   -7.02  &    3.90  \\
  15 &          (A+B{\it t}) ln({\it t}) &  2 & 3.440$\cdot 10^{0}$ &-1.447$\cdot 10^{-3}$ &          &  10.513 & 18  &    0.58  &   -8.16  &    3.54  &   -6.87  &    4.05  \\
  16 &  (A+B{\it t}+C{\it t}$^2$) ln({\it t}) &  3 & 1.496$\cdot 10^1$ &-1.345$\cdot 10^{-2}$ & 3.125$\cdot 10^{-6}$ &  10.262 & 17  &    0.60  &   -5.85  &    5.85  &   -4.36  &    6.56  \\
\hline
\end{tabular}
}
\tablefoot{A,B, and C are the values of the optimized parameters for each model.
The quoted number of degrees of freedom (d.o.f.) is equal to the number of data points, $n$, minus the number of independent fitting parameters, $p$. 
$\Delta$AIC ($\Delta$BIC) is the pairwise difference with respect to the value of the AIC (BIC) of the most likely model.
See text for details.}
\end{table*}

\section{GCR modulation and \ti\ Activity in Meteorites}

The propagation of GCRs through the heliosphere is completely described by the transport equation derived by Parker (1965). 
A simpler force-field approximation (Gleeson \& Axford 1968; Caballero-Lopez \& Moraal 2004) is often used in the literature in the quasi-steady state assumption of negligible adiabatic energy loss rate, no drift and no additional sources of GCRs within the heliosphere.
This approximation provides a useful parameterization of the modulation of the GCR flux, which can be described in terms of a quasi-steady state balance between the radially inward diffusion through the HMF irregularities and the radially outward convection with the solar wind speed.
In this approximation, Parker's transport equation is reduced to a simple convection-diffusion equation that depends on a modulation parameter $\phi$, describing the modification of the local interstellar spectrum (LIS) in the heliosphere as a function of time $t$ and heliospheric distance $r$
\begin{equation}
 \phi(t,r) = \int_r^{r_b} \frac{v_{\rm sw}(t,r')}{3\kappa(t,r')}dr',
\end{equation}
where $\kappa$ is the rigidity-independent diffusion coefficient, $v_{\rm sw}$ is the solar wind speed, and $r_b$ is the outer boundary of the heliosphere.
Since the heliospheric modulation is ultimately defined by the solar activity, the time-dependent parameter $\phi$ can be considered as a proxy of past solar global magnetic activity.
Despite its simplicity, the modulation parameter $\phi$, usually given in units of MV, is able to qualitatively describe the modulation of the LIS at 1 AU during the spacecraft era (Caballero-Lopez \& Moraal 2004). 
Neutron monitor and satellite data show that $\phi$ increases with solar activity.
While typical monthly average values within a solar cycle have ranged from 400 MV to 1100 MV over the modern era, decadal averages of solar cycles typically yield $\phi \sim 550-700$ MV during grand solar maxima and $\phi \sim 100-200$ MV during grand solar minima (McCracken et al. 2004).
The production rate (or activity $A(t)$ as a function of time) of \ti\ in meteorites depends on the GCR intensity in space.
After the fall of the meteorites, this activity can be measured in laboratory since the radioactive decay chain \ti\ $\to$ \sca\ $\to$ \ca\ leads to the emission of $\gamma$-rays that are detectable with well designed $\gamma$-ray spectrometers. 

Figure 1 shows the \ti\ activity measured in 20 stony meteorites which fell between 1766 and 2001 (Taricco et al. 2006, 2008, 2016).
These measurements were performed at the Monte dei Cappuccini Laboratory (INAF--OATo) in Turin, Italy, using $\gamma$-ray spectrometers (HPGe+NaI) described in detail in Taricco et al. (2006) and Colombetti et al. (2013).
The \ti\ activity values have been normalized for shielding of the meteorite fragment in space as well as corrected for the concentration of the Fe and Ni target elements.
The sources of error are due to uncertainties in concentrations of K, Fe and Ni ($<3$\% in most meteorites) and shielding corrections ($<15$\%) because the \ti\ isotope production depends on the size of the meteoroid and the shielding depth. 
The expected \ti\ activity at a given time $t$ is
\begin{equation}
A(t) =  {{f} \over {\tau}} \int_{-\infty}^{t}  Q(t')e^{-(t-t') /\tau}  dt'
\end{equation}
where $Q(t)$ is the \ti\ production rate, $\tau = 85.4 \pm 0.9$ yr its mean life time, and $f=3.4$ is a scaling factor (Usoskin et al. 2006) corresponding to the chemical composition and pre-atmospheric size of the Torino meteorite (Bhandari et al. 1989), which was adopted as a reference meteorite.
The expected \ti\ production rate $Q$ in a stony meteorite as a function of the modulation parameter $\phi$, was calculated by Michel \& Neumann (1998) for discrete values of $\phi$ ranging from 300 to 900 MeV and can be aptly modeled by the expression (Usoskin et al. 2006)
\begin{equation}
Q(\phi) = 0.82 + 2.07\exp(-0.00319\phi),
\end{equation}
with $Q(\phi)$ in units of [dpm kg$^{-1}$].
From a formal point of view, Eq. (2) is equivalent to an integral equation with infinite lower limit of integration whose exponential kernel only depends on the difference of the arguments. 
Given $A(t)$, it is possible to solve Eq. (2) as (e.g., Polyanin \& Manzhirov 1998) 
\begin{equation}
Q(t) = \frac{\tau}{f}\left(\frac{dA(t)}{dt} + \frac{1}{\tau}A(t)\right)
\end{equation}
to infer the unknown $Q(t)$ dependence and, by inverting Eq. (3), recover the temporal evolution of $\phi$, which will be directly related to the HMF strength.

Since our dataset consists of 20 discrete data points (see Fig. 1), a simple analytical function $A(t)$, representing the long-term evolution (or trend) of the observed activity, can be taken as the required input to solve Eq. (4).
In general, the trend of a time series can be considered as a smooth additive component containing information about its global change.
A common mathematical approach to the problem of trend extraction is given by fitting a low-order polynomial (usually a linear or quadratic one) to the data. 
By adopting a simple comparison of the maximum likelihood of different analytic models, the model with the most parameters is deemed to be favored, although not all parameters might actually be relevant. 
In deciding which specific model is the best, criteria are therefore needed allowing model comparisons to avoid both simplistic (underfitting) and overly complex (overfitting) models.
Although minimizing the $\chi^2$ is a typical approach for finding the best fit parameters for a certain model, it is however insufficient for deciding whether the model itself is better supported by the data because it does not take into account the relative structural complexity of the models.
In this latter circumstance, it is now customary to use alternative methods based on information theory rather than hypothesis testing.
The use of information criteria (ICs) that extend the usual $\chi^2$ criterion is now common in several fields (see, e.g., Liddle 2007 for an application to cosmology) where it is needed to compare the evidence for and against competing models.
IC methods provide the relative ranks of two or more competing non-nested models, also yielding a numerical measure of confidence (analogous to likelihoods or posterior probabilities in traditional statistical inference) that each model is the most likely.
Two popular tools for optimal model selection are the Akaike information criterion (AIC; Akaike 1973), defined as ${\rm AIC} =  -2\ln\mathcal{L}_{\rm max} + 2p(p-1) /(n-p-1)$
and the Bayesian information criterion (BIC; Schwarz 1978), defined as   ${\rm BIC} =  -2\ln\mathcal{L}_{\rm max} + p\ln{n} $, where $\mathcal{L}_{\rm max}$ is the maximum likelihood, $p$ the number of parameters of the model and $n$ is the number of data points used in the fit.
For normally distributed errors, $-2\ln\mathcal{L}_{\rm max} = n\ln({\chi^2_{\rm min}/n}) $  (Burnham \& Anderson 2002).
AIC focusses on the Kullback–Leibler information entropy (Kullback \& Leibler 1951) as a measure of information lost when a particular model is used in place of the (unknown) true model. 
BIC comes from approximating the Bayes factor (Jeffreys 1961), which gives the posterior odds of one model against another, presuming that the models are equally favored prior to data fitting.
Both criteria balance simplicity (measured by the dimension of the fitted model parameter space, $p$) and goodness of fit (measured by the dimension of the log-likelihood $\mathcal{L}$ and can be used for both linear and non-linear models.
If there are several competing models for the data, the one with the lowest index value is assessed as the one most likely to be nearest to the unknown model that generated the data. 
In order to assess how much statistical importance we should attach to a difference in the AIC values between the best model and the other candidate models, we have also computed the pairwise differences $\Delta$AIC = AIC - AIC$_{\rm min}$, where AIC$_{\rm min}$ is the AIC score of the most likely model. 

Table 1 shows the results obtained from fitting 16 different analytic models to our data.
Due to the limited number of degrees of freedom, our search was limited to analytical functions with less than four free parameters. 
The optimized parameters for each model, A, B, and C, are listed together with the best-fit $\chi^2$ values for each model, the number of degrees of freedom (d.o.f.) and the $\chi^2$ per d.o.f. $\chi^2_\nu$.
All models have a reduced $\chi^2_\nu$ between 0.52 and 0.62 and are compatible with the measurements at an 88\% confidence level or better.
Model 11 is, however, optimal with respect to all listed indexes.
The probability $P$ that this model is to be preferred over an alternative candidate model, given by $P = {\rm e}^{-\Delta {\rm AIC}/2}/(1+{\rm e}^{-\Delta {\rm AIC}/2})$  (e.g., Burnham \& Anderson 2002),
yields values ranging from 77 to 96\%, implying that all other competing models are disfavored (according to the above statistical standards) as compared to Model 11.
The BIC index yields similar results, pointing to the fact that Model 11 is indeed the most likely to be correct in the frame of information theory.
Finally, since the two models with the lowest $\chi^2_\nu$, AIC and BIC scores (models 11 and 12) are nested (one model is a simpler case of the other), it is also possible to compute a traditional statistical hypothesis testing such as an F-test and choose a model according to the F-ratio.
From the latter, a P value of 0.81 is calculated, implying that the simpler (null) model cannot be statistically rejected.
Therefore, Model 11, represented by the analytical function $y(t)=t/({\rm A}+t)$ with best-fit parameter given by ${\rm A} = -1526.2^{+10.7}_{-10.1}$ was selected as the best model for the trend of the \ti\ activity for use in further analyses.
Figure 1 shows the above trend, superimposed onto the measured data, together with the $1\sigma$  and $2\sigma$ confidence bands obtained by imposing $\chi^2 \leq \chi^2_{min} + \Delta\chi^2$ with, respectively,  $\Delta\chi^2=1.0$ and   $\Delta\chi^2=4.0$ (e.g., Press et al. 1992).
In the same plot, we show, as an inset, the time profile of the modulation potential $\phi$ (in MV) obtained by inverting Eq. (3). 
We would like to emphasize that, strictly speaking, measurements of \ti\ activity in meteorites do not allow us to perform a true reconstruction of the HMF because of the time integration. 
What is estimated in this work is not the true temporal variability of the HMF but the secular trend given by a prescribed functional shape chosen by means of statistical and information criteria. 
A reliable reconstruction would require the use of the derivatives of the measured activity, but this is impossible for such sparse and noisy data as those shown in Fig. 1. 

\begin{figure}
\centering
\includegraphics[width=9.3cm]{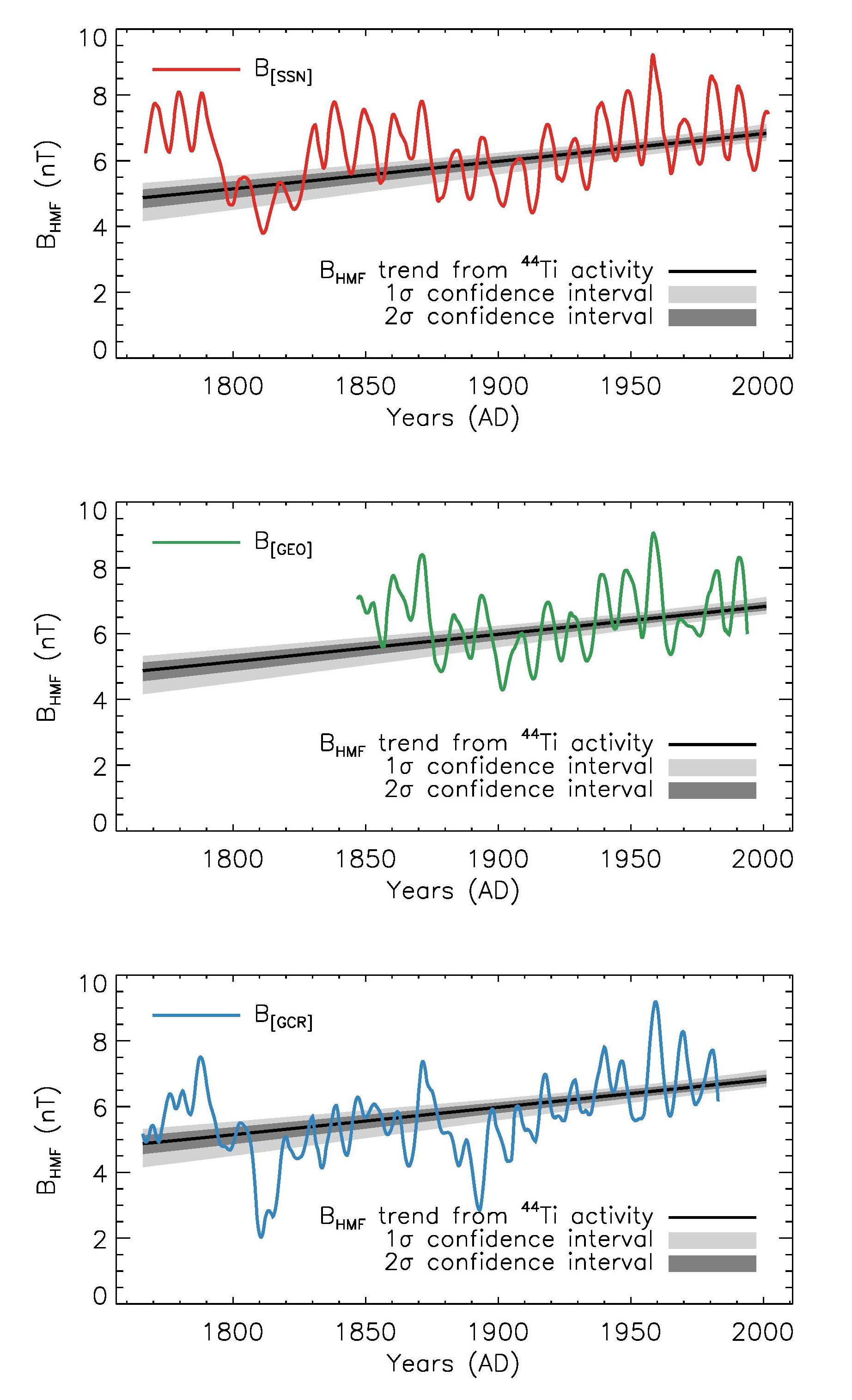}
\caption{
Long-term evolution of $B_{\rm HMF}$ between 1766 and 2001 ({\it black} line) as obtained by using the $A(t)$ trend derived from the measured \ti\ data. 
The 1$\sigma$ (68.27\%) and 2$\sigma$ (95.45\%) confidence bands (depicted with dark-gray and light-gray shaded areas, respectively) were obtained by propagating the uncertainties in the model (see text for details). 
The coloured curves are the sunspot-based ({\it top}), geomagnetic-based ({\it middle}) and \bex-based ({\it bottom}) composite reconstructions of the $B_{\rm HMF}$ after passing through a 1,4,6,4,1 binomial filter as given by Owens et al. (2016a,b).}
\end{figure}

\section{Long-term evolution of the HMF intensity}

Given the functional form for $A(t)$, the modulation parameter $\phi(t)$ can be thus obtained, through Eq. (4), by inverting Eq. (3).
We point out that since meteorites usually originate from the asteroid belt at $\sim 2-3$ AU, where the GCR modulation is slightly weaker than at 1 AU, $\phi(t)$ was reduced by 1\% to account for the radial dependence of the GCR flux in the heliosphere (Caballero-Lopez \& Moraal 2004; Usoskin et al. 2006).
Provided that $v_{\rm sw}$ and $\kappa$ in Eq. (1) are spatially independent and that $\kappa \propto B^{-\alpha}$ (Caballero-Lopez et al. 2004), the modulation potential $\phi$ yields the variation of the product $v_{\rm sw} B_{\rm HMF}$ with time.
Assuming constant solar wind speed, which is valid on long-term scales, the HMF strength $B_{\rm HMF}$ can be related to the modulation potential $\phi$ according to 
$B_{\rm HMF}(t) = B_{\rm HMF,0} \left[\phi(t)/\phi_0 \right]^{1/\alpha}$ (e.g., Steinhilber et al. 2010), where the constants are normalization factors obtained by fitting data at 1 AU. 
Experimental data imply $\alpha = 1.9 \pm 0.3$, $\phi_0 = 615$ MV, and $B_{\rm HMF,0} = 6.6$ nT (Steinhilber et al. 2010). 
In the calculation of the 1$\sigma$ (68.27\%) and 2$\sigma$ (95.45\%) confidence intervals for $B_{\rm HMF}$, we also propagated the uncertainty in $\tau$ ($\Delta\tau \approx 0.9$) and the uncertainty in the relation expressing the expected \ti\ production rate $Q$ in a stony meteorite as a function of the modulation parameter $\phi$ (Eq. 3) estimated as in Usoskin (2006).
We remark that the above relation (Eq. 3) has been obtained assuming the LIS of Castagnoli \& Lal (1980) in the calculation of the modulation parameter $\phi$ and is therefore consistent with the parameterization used in Steinhilber et al. (2010).
Here it is assumed that the exponent $\alpha$ has not changed in time due to its time invariant nature. 
$\alpha$ will be time invariant if the nature of the diffusion processes due to the turbulence of the solar wind are time-independent (Caballero-Lopez et al. 2004; McCracken 2007).
The above equation is even valid for any spatial dependence of the variables, provided that this spatial dependence does not change with time.

Figure 2 shows the long-term evolution of $B_{\rm HMF}$ between 1766 and 2001 ({\it black} line) as obtained by using the $A(t)$ trend derived in the previous section.
The 1 and 2 $\sigma$ confidence bands (depicted with dark-gray and light-gray shaded areas, respectively) were obtained by propagating the errors according to the uncertainty included in the functional representation for the \ti\ activity trend, including the uncertainty in $\alpha$, $\tau$ and in Eq.(3).   
Uncertainties were propagated through all calculations using a Monte-Carlo approach.
The trend is found to be linear and given by a $B_{\rm HMF}(t) = -9.879+0.00835 t$ with $t$ in years.
We find that the HMF strength has varied from $4.87^{+0.24}_{-0.30}$ nT in 1766 to $6.83^{+0.13}_{-0.11}$ nT in 2001, thus implying an overall average increment of $1.96^{+0.43}_{-0.35}$ nT over 235 years since 1766 reflecting the modern Grand maximum. 

The long-term evolution of $B_{\rm HMF}$ obtained with the measured \ti\ activities was compared with the most recent state-of-the-art reconstructions of the near-Earth heliospheric magnetic field strength $B_{\rm HMF}$ obtained by Owens et al. (2016a,b) based on geomagnetic, sunspot number and cosmogenic isotope data. 
To compute $B_{\rm HMF}$ from sunspot records, Owens et al. (2016a) used two different reconstruction methods. 
The first, a purely empirical approach that assumes $B$ to be linearly correlated to the square root of sunspot number (e.g., Svalgaard \& Cliver 2005; Wang et al. 2005), makes no explicit assumptions about the physical mechanisms by which $B_{\rm HMF}$ varies.
The second derives $B_{\rm HMF}$ from an estimate of the open solar flux, which is modeled as a continuity equation, where the flux source rate is assumed to vary in phase with sunspot number (Solanki et al. 2000; Vieira \& Solanki 2010). 
From the above two models, considering four corrected sunspot records, they finally derived a weighted composite time series,  $B_{\rm [SSN]}$, extending from 1750 to 2013 (Fig 2a).
Owens et al. (2016a) also evaluated two complementary estimation procedures based on geomagnetic data (e.g., Lockwood et al. 2013; Svalgaard 2014) that rely on the interdiurnal variation observed at geomagnetic observatories since 1845 but differ in the number of geomagnetic observatories and data used.
From the above two models, the authors derived a weighted composite time series, $B_{\rm [GEO]}$, extending from 1845 to 2013 (Fig 2b).
Finally, Owens et al. (2016b) presented two independent cosmogenic radionuclide reconstructions derived from the annual measurements of the \bex\ concentrations in the Dye 3 (Greenland) and the North Greenland Ice Core Project (North GRIP) ice cores, based on the works of McCracken \& Beer (2015) and Usoskin et al. (2015).
An average composite time series, $B_{\rm [GCR]}$, was obtained by averaging together these two independent cosmogenic radionuclide reconstructions in the common intervals from 1766 to 1982 (Fig. 2c).

In Fig. 2, we show the long-term evolution of $B_{\rm HMF}$ obtained in this work along with the above three average reconstructions, passed through a 1,4,6,4,1 binomial filter (Aubury \& Luk 1996) to remove the high-amplitude variations due to the 20\% standard deviation variability of the annual \bex\ data (see Owens et al. 2016b).
In Fig. 3, for a better comparison, the same trend is superimposed to linear fits to the above three average reconstructions.
The long-term evolution of $B_{\rm HMF}$ obtained with the measured \ti\ activities is clearly significantly lower (at least before about 1940) than the similar linear fits obtained from the $B_{\rm [GEO]}$ and $B_{\rm [SSN]}$ reconstructions.
In particular, the slope of the linear fit to the $B_{\rm [SSN]}$ time series would imply an overall increment of only about 0.9 nT between 1766 and 2001 as compared to an overall increment of $1.96^{+0.43}_{-0.35}$ nT as evinced from our measurements. 
Instead, an excellent agreement, well within the $1\sigma$ confidence interval, is found between the trend derived in this work and the linear fit of the composite $B_{\rm [GCR]}$ (bottom panel of Fig. 2). 
This last result might be apparently unsurprising in the light of the fact that both trend reconstructions have been obtained through the analysis of cosmogenic radionuclide data, albeit with a completely different approach. 
However, we point out that our results are not influenced by terrestrial effects, such as the variable geomagnetic field, climatic changes or deposition rate variations that affect all other models. 
We also point out that the Suess effect and the bomb-effect do not allow to use \cxiv\ data after the second half of the 19th century.
As such, the outcome of this work, based on the measurement of the cosmogenic \ti\ activity detected in meteorites, supports the validity of the reconstructions obtained from the analysis of cosmogenic radionuclide (e.g., \bex) data in the terrestrial reservoirs and might be used as an independent check for future reconstructions.

\begin{figure}
\centering
\includegraphics[width=9.3cm]{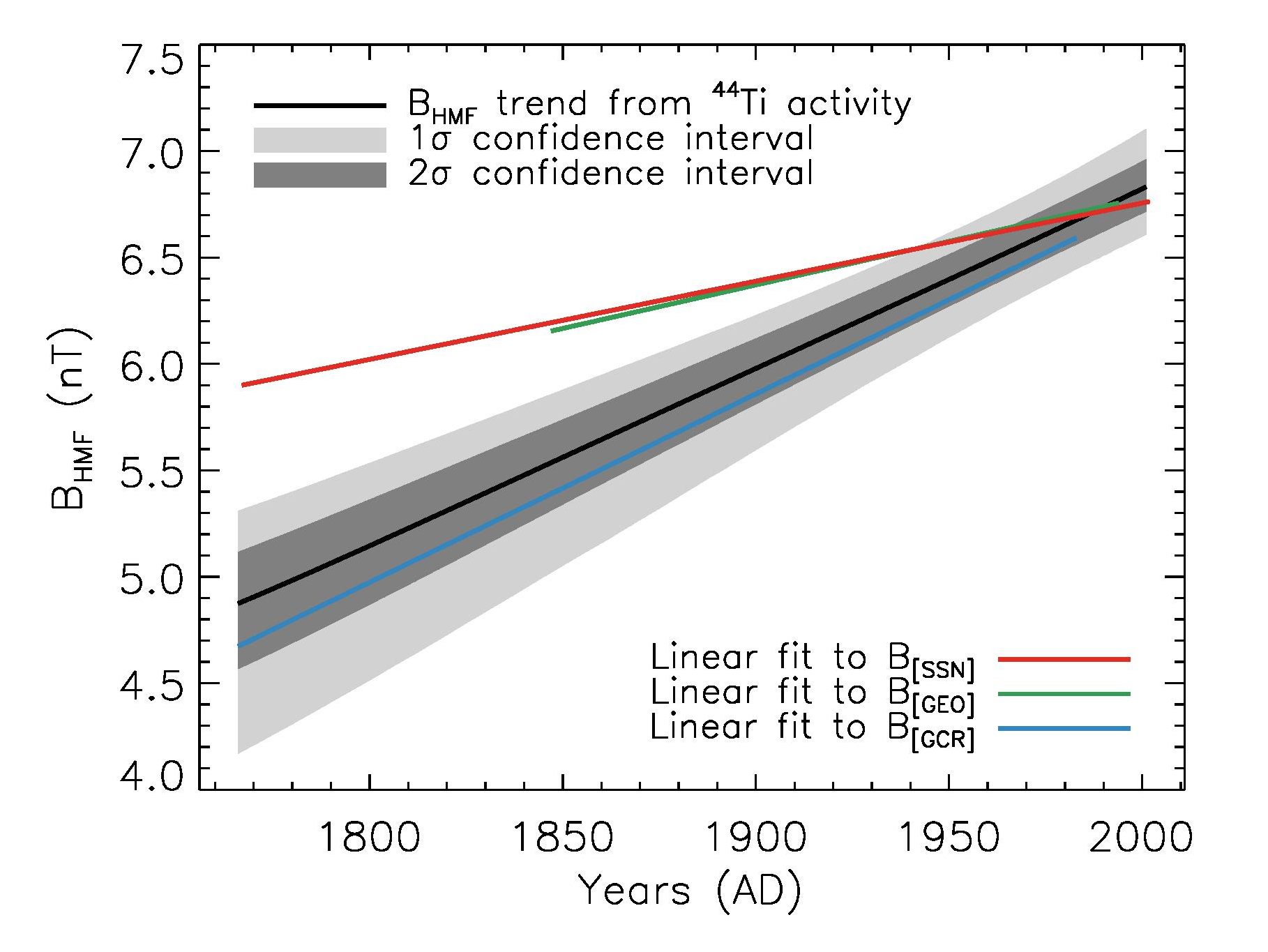}
\caption{
Long-term trend of $B_{\rm HMF}$ between 1766 and 2001 ({\it black} line) as obtained by using the $A(t)$ trend derived from the measured \ti\ data. 
The 1$\sigma$ (68.27\%) and 2$\sigma$ (95.45\%) confidence bands (depicted with dark-gray and light-gray shaded areas, respectively) were obtained by propagating the uncertainties in the model (see text for details). 
The colored curves are linear fits to the geomagnetic-based, sunspot-based and \bex-based composite reconstructions of the $B_{\rm HMF}$ as given by Owens et al. (2016a,b).
}
\end{figure}

\section{Discussion and conclusions}

Historical records of sunspots, geomagnetic indices and aurorae suggest that the HMF has changed substantially over the past centuries. 
Several attempts have been made to reconstruct the HMF temporal dependence over the last centuries.
There are, however, important discrepancies among the results obtained by the various authors using different proxies of solar magnetic activity and different methods.
One major factor affecting the proposed HMF reconstructions is the influence of secular variations of the geomagnetic field on aurorae, geomagnetic indices and production rates of cosmogenic nuclides  (e.g., \bex\ in ice sheets and \cxiv\ in tree trunks) generated by nuclear reactions of GCRs incident on the Earth's atmosphere. 
Another important factor is the imprecisely known contribution of the terrestrial processes leading to the redistribution of these isotopes in the terrestrial archives including fossil fuel burning and bomb effect.
As for forward models that rely on the properties of active regions (sunspots, faculae) and the historical sunspot record to infer the HMF near Earth (e.g., Solanki et al. 2000, 2002; Wang \& Sheeley 2003), they all involve ad hoc factors that remain to be as yet firmly established.

The unique approach proposed in this paper, based on the measurement of the cosmogenic \ti\ activity detected in meteorites, is designed to overcome most of the problematics that affected the previous efforts, thus yielding a powerful independent tool to reconstruct the temporal dependence of the long-term evolution of the HMF through the last two and a half centuries.
Within the given uncertainties, our result is compatible with a HMF increase from $4.87^{+0.24}_{-0.30}$ nT in 1766 to $6.83^{+0.13}_{-0.11}$ nT in 2001, thus implying an overall average increment of $1.96^{+0.43}_{-0.35}$ nT over 235 years since 1766 reflecting the modern Grand maximum.
When compared with the trend derived from the most recent HMF reconstructions given by Owens et al. (2016a,b), the long-term variation obtained in this work is in agreement with the average \bex-based reconstruction within a $1\sigma$ confidence interval. 
The sunspot-based and geomagnetic-based reconstructions appear to underestimate the slope of the overall trend when compared with the result obtained in this work, resulting flatter than evinced from our data.
Our result is also concordant with the one obtained by Asvestari et al. (2017) who tested different sunspot number series against the same record of cosmogenic \ti\ activity measured in meteorites used in this work.

New measurements of the \ti\ activity in meteorites that have been recently acquired by our group are ongoing through a much improved $\gamma$-ray spectrometer. 
These data will allow to further thicken and extend the investigated period by more than one decade, while possibly reducing the uncertainties in the HMF reconstruction.

\begin{acknowledgements} 
The authors would like to thank the anonymous referee whose suggestions allowed us to significantly improve this paper.
We also thank Dr. P. Janardhan and Dr. A. C. Das for careful reading of the manuscript and helpful discussions.
\end{acknowledgements}

\end{document}